\documentclass[prb,twocolumn,showpacs,floatfix]{revtex4}

\usepackage{graphicx}%
\usepackage{dcolumn,float}
\newcommand{\no}{\noindent}

\newcommand{\beq}{\begin{equation}}
\newcommand{\eeq}{\end{equation}}
\newcommand{\qq}{\begin{eqnarray}}
\newcommand{\qqq}{\end{eqnarray}}

\newcommand{\ve}{\varepsilon}

\begin{document}

\title{What are the elementary excitations
of the BCS model in the canonical ensemble?}

\date{\today}

\author{Jos\'e Mar\'{\i}a Rom\'an$^{1}$, Germ\'an\ Sierra$^{1}$, and
 Jorge\ Dukelsky$^{2}$ } 

\affiliation{ 
$^{1}$Instituto de F\'{\i}sica Te\'orica, CSIC/UAM,
Madrid, Spain.\\
$^{2}$Instituto de Estructura de la Materia, CSIC, Madrid, Spain
}

\begin{abstract}
We have found the elementary excitations of the exactly solvable
BCS model for a fixed number of particles.  These turn out to
have a peculiar dispersion relation, some of them with no 
counterpart in the Bogoliubov picture, and unusual counting
properties related to an old conjecture made by Gaudin.
We give an algorithm to count the number of excitations
for each excited state and a graphical interpretation 
in terms of paths and Young diagrams. For large systems
the excitations are described by an effective Gaudin model,
which accounts for the finite size corrections to BCS.  
\end{abstract}


\maketitle

\newcommand{\bb}{\boldsymbol{\beta}}
\newcommand{\ba}{\boldsymbol{\alpha}}


\section{Introduction}

The paradigmatic model to study the superconducting
properties of metals \cite{BCS} and nuclei \cite{nuclear}  
is the pairing model
proposed by Bardeen, Cooper and Schrieffer. 
The ground state (GS) and excitations of the BCS 
model are well known in the grand canonical 
(g.c.)\ ensemble, and explain the behavior of systems with
large number of particles. However for small systems, such
as nuclei or nanograins, one is forced to work
with a fixed number of particles, where
the g.c.\ BCS wave function, including its
projected version, are 
not adecuate. The problem is due to the strong
pairing fluctuations, which a mean field approach 
cannot deal with properly. An alternative approach
is provided by exact numerical methods, as 
the DMRG \cite{DMRG}, but their 
complexity somehow obscures
the Physics behind. Fortunately enough the reduced
BCS model, characterized by a unique pairing coupling $g$,
is exactly solvable, as was shown 
by Richardson \cite{exact}. 
This exact solution has been recently
used in connection with superconducting 
nanograins \cite{random}
(see Ref.~\onlinecite{vDR} for a review). 

Most of the exact studies deal with the GS and the 
excited states that are obtained by breaking Cooper pairs. 
However one must also consider the promotion of pairs to 
higher energy levels (bosonic pair-hole excitations).  
This paper focus on the later type of excitations since 
the former ones can be easily included into our formalism. 
We shall indicate the peculiar dynamics and the unusual
counting properties exhibited by the excitations of the 
exactly solvable BCS model, some of them with no analogue in the
standard picture of Bogoliubov quasiparticles.
These features account for the exact finite size corrections 
to the thermodynamic limit, obtained from the standard
 BCS treatment.

In Section~\ref{Richardson} we introduce the superconducting
system described by Richardson and Gaudin models, and  
we present here our conjecture about what represents
their elementary exciations.
Section~\ref{States} is devoted to classify the
excited states according to what will be interpreted
as elementary excitations in Section~\ref{Large-N}.  
In this respect a diagramatic representation of the 
excited states turns out to be very useful.  
The thermodynamic limit of our theory is presented in 
Section~\ref{Large-N}, which confirms our conjecture
about the elementary excitations.  A comparison with the BCS
theory is also presented.  
Finally, Section~\ref{Conclusions} contains our conclusions.


\section{Richardson and Gaudin models}
\label{Richardson}

Let us consider a fermionic system with 
$N$ single particle energy levels. 
The reduced BCS Hamiltonian decouples
the levels which are singly occupied
and one is left with those
that are either empty, $|0\rangle$ (hole)  
or doubly occupied, 
$b^\dagger_j |0 \rangle$ (pair) 
with an energy $\ve_j$ 
($b^\dagger_j = c^\dagger_{j,+} c^\dagger_{j,-}$ 
is a hard core boson operator that creates a pair 
of two time-reversal states).
We shall suppose that the singly occupied levels 
have been removed. Since the latter decouple, their effect 
can be considered easily by adding their free
energy $\ve_j$ to the total energy.
The complete system will be treated elsewhere. 
The reduced BCS Hamiltonian reads, 
\beq
H_{BCS} = \sum_{j=1}^N   \varepsilon_j b^\dagger_j b_j - G \,
\sum_{j,j'=1}^N \; b_j^\dagger b_{j^{\prime}} \;, 
\label{1}
\eeq
\no where $G$ is a dimensionful coupling
constant. The standard model employed to study
nanograins is given by the choice
$\ve_j = d (2 j - N -1)$,
where $d=\omega/N$ is the single particle energy 
level spacing and $\omega/2$ is the Debye energy \cite{vDR}.
The coupling $G$ can be written as  
$G= g d $, where $g$ is dimensionless. 
The value of the bulk BCS gap, $\Delta_{BCS}$,  
of the equally space model 
is given by $\Delta_{BCS} = \Delta/2$,
where $\Delta = \omega/\sinh(1/g)$. 

The eigenstates of (\ref{1}) with $M$ pairs 
are given by \cite{exact}
\beq
|\{ E_\mu \}_{\mu=1}^M 
\rangle = \prod_{\mu = 1}^M B_\mu^\dagger |0 \rangle, 
\qquad
B_\mu^\dagger = \sum_{j=1}^N \frac{ b^\dagger_j}{\varepsilon_j - E_\mu},
\label{2} 
\eeq
\no where the parameters $\{ E_\mu \}_{\mu =1}^M$
satisfy the Richardson equations
\beq
\frac{1}{G } 
= \sum_{j=1}^N \frac{1}{ \varepsilon_j - E_\nu}
- \sum_{\mu \neq \nu}^{M} \frac{2}{ E_\mu - E_\nu}\;, 
\; \nu =1, \dots, M.  
\label{3}
\eeq
\no The total energy of the state (\ref{2}) reads
$E= \sum_{\mu=1}^M \; E_\mu$. 
The number of solutions of Richardson 
Eqs.~(\ref{3}) is given by the binomial coefficient
$C^N_M = {N\choose M}$, 
and coincides with the dimension of the Hilbert space,
${\cal H}^N_M$, of states with $M$ pairs distributed
into $N$ different levels. 
Then it  is natural to label Richardson states by 
a set of integers \mbox{$I = \{j_1, \dots, j_M \}$}
corresponding to the value that the pair energies 
$E_\mu$ take at $g = 0$, i.e.\ some of the $\ve_j$'s.

The BCS model can be mapped into a spin system
which at $g \rightarrow \infty$ has $SU(2)$ 
symmetry. Based on this fact  
Gaudin \cite{G-book} made the conjecture that given a
solution $\{ E_\mu(g) \}_{\mu=1}^M$  of Eqs.~(\ref{3}), 
and taking the limit  $g \rightarrow \infty$, 
a subset of them, say $\{ E_\alpha(\infty)\}_{\alpha=1}^{N_G}$,
remain finite and satisfy the equations 
\beq
0 
= \sum_{j=1}^N \frac{1}{ \varepsilon_j - E_\alpha}
- \sum_{\beta \neq \alpha}^{N_G} \frac{2}{ E_\beta - E_\alpha}\;, 
\ \alpha =1, \dots, N_G, 
\label{4}
\eeq
\no while the
remaining $M-N_G$ roots tend to infinity and 
satisfy Eqs.~(\ref{3}) with all $\ve_j$'s set to zero. 
The number $N_G$ of finite roots takes values from 0 to $M$.
The number of  solutions of Eqs.~(\ref{4})
is given by $d_{N_G} = C^N_{N_G} - C^N_{N_G-1}$ \cite{G-book}. 
Therefore in the large $g$ limit the $C^N_M$ 
Richardson's solutions would be classified in terms of $N_G$
according to Table~\ref{table1}. 
Consistency is guaranteed by the equation 
$C^N_M = \sum_{N_G=0}^M  d_{N_G}$.


\begin{table}[H]
\begin{center}
\begin{tabular}{|c|c|c|c|c|c|}
\hline
\# of solutions & $d_M$ & $d_{M-1}$ & $\cdots$ & $d_1$ & $d_0=1$ \\
\hline
$E_\mu$ finite & $M$ & $M-1$ & $\cdots$ & 1 & 0   \\
$E_\mu$ infinite & $0$ & $1$ & $\cdots$ & $M-1$ & $M$ \\   
\hline
\end{tabular}
\end{center}
\caption{Classification of roots in the $g \rightarrow \infty$ limit.}
\label{table1}
\end{table}



\begin{figure}[t]
\begin{center}
\includegraphics[width= 8 cm,angle= 0]{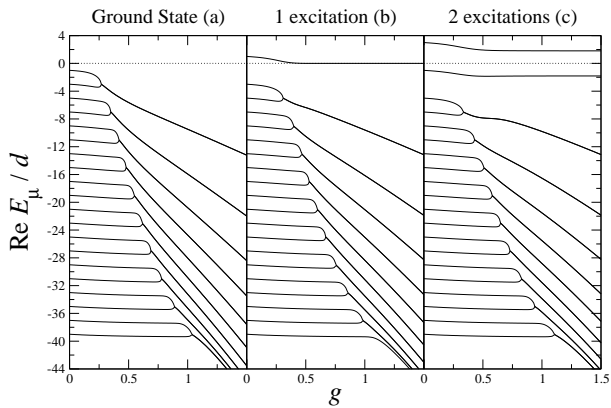}
\end{center}
\caption{Real part of  $E_\mu$ 
for the equally space model 
with $M=N/2=20$ pairs and $N_G =0,1,2$ excitations. 
}
\label{fig1}
\end{figure}


We show in this paper that Gaudin finite energies represent
the elementary excitations of the superconducting system in
the canonical ensemble.  Their peculiar dispersion relations
and the unusual counting properties will account for the
finite size corrections to the mean-field BCS treatement
of superconductivity.

This result is motivated by the excitation
energy for large values of $g$, namely,
$E_{exc} \equiv E - E_{GS} \sim g \omega N_G [1 - (N_G-1)/N]$,
and the gap $\Delta \sim g \omega $.
Thus, in the large $N$ limit the excitation
energy goes as $E_{exc} \sim N_G \Delta$.
Which allows us to think of the state as a set of $N_G$
elementary excitations contributing each with an energy
$\Delta$ to the total energy.

In Section~\ref{Large-N} we extend this result for the
whole range of $g$. In the meanwhile next Section is devoted to the proof of
Gaudin's conjecture given in Table~\ref{table1}.
We also obtain a formula which gives the number of finite 
Gaudin energies $N_G$ for a given Richardson configuration $I$,
and therefore the number of elementary excitations.


\section{Classification of excited states}
\label{States}

\subsection{Simple examples}

\begin{figure}[t]
\begin{center}
\includegraphics[width= 8 cm,angle= 0]{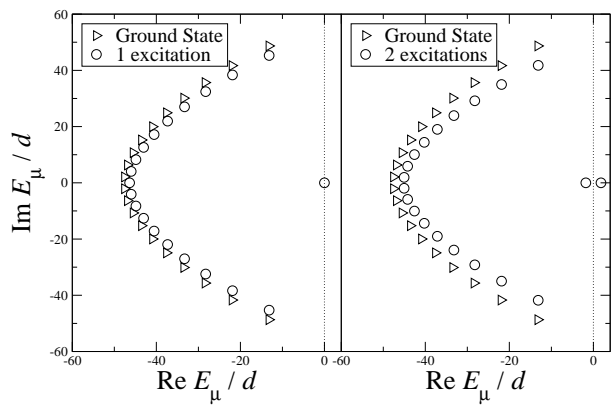}
\end{center}
\caption{Position of the $M=20$ pairs of the states
of fig.~\ref{fig1} at $g=1.5$. The arcs $\Gamma_{I_1}$
(19 pairs) and   $\Gamma_{I_2}$ (18 pairs) 
are a slight modification of the GS arc $\Gamma_{I_0}$
(20 pairs). 
} 
\label{fig5}
\end{figure}

Let us first consider the simplest examples given
by the excited states with one and two 
energies remaining finite, i.e.\ $N_G = 1$ and $2$.
Representatives of these, together with the GS
are shown in Fig.~\ref{fig1}, which depicts
to the real part of the energies,
and in Fig.~\ref{fig5}, which shows the distribution
of the energies in the complex plane for $g = 1.5$
and a system with \mbox{$M=20$} pairs
at half filling, i.e.\ $N=2M$.  
As a general feature we see that
for small $g$ all parameters $E_\mu$ 
are real, and as $g$ grows some of them
collapse and become complex conjugate 
pairs, which share their real part
(this corresponds to two curves merging
into a single one in Fig.~\ref{fig1}). 
Fig.~\ref{fig5} shows how the energies
$E_\mu$ arrange themselves into an arc 
which opens up to infinity as $g \to \infty$.


The state of Fig.~\ref{fig1}a corresponds to the GS of the system,
and it is labelled by
\mbox{$I_0 = \{ 1,2, \dots, M \}$}, which at $g=0$
is identical to the Fermi state (FS). 
As \mbox{$g \rightarrow \infty$} all the roots
become complex and scape to infinity.
According to Table~\ref{table1},
this is the only state where this may happen, hence
$N_G(I_0) = 0$. 

The lowest excited state \mbox{$I_1 = \{ 1, \dots, M-1,M+1 \}$}
is shown in Fig.~\ref{fig1}b.  
The last root $E_M$, which is equal to $\ve_{M+1}$
at $g=0$, stays finite as $g \rightarrow \infty$, while
the remaining $M-1$ roots go to infinity, thus
$N_G(I_1) = 1$. 
All the states with $N_G=1$ can be obtained
from the FS by: \mbox{1) promoting} the nearest 
pair below the Fermi level (FL) into one
of the $N-M$ empty levels above it,  
or \mbox{2) moving} the 
nearest hole above the FL into one of the $M$
occupied levels below it. The state $I_1$
can be obtained in both ways. Hence
the number of $N_G=1$ excited states
is $d_1 = (N-M) + M -1$. 

The state of Fig.~\ref{fig1}c, 
\mbox{$I_2 = \{1,\dots, M-2,M,M+2 \}$}, has $N_G(I_2)=2$.
All the states  with $N_G=2$ 
can be obtained in three different ways from the FS:  
\mbox{1) moving} the two pairs just below the FL
into the $N-M$ empty levels ($C^{N-M}_2$ states),
\mbox{2) moving} the two holes just above the FL 
into the $M$ occupied levels ($C^M_2$ states), 
or \mbox{3) moving} one of the $M-1$ pairs in the FS,
except the closest to the FL, into one of the
$N-M-1$ vacancies above the FL, except 
the closest to the FL ($(N-M-1)(M-1)$ states). 
The state with two holes just below and two
pairs just above the FL is generated by the 
rules 1 and 2, thus the number of states  
is the expected one,  
$d_2 = C^{N-M}_2 + C^M_2 + (N-M-1) (M-1) -1$. 
This example shows that the value of $N_G$ for
a generic state depends  dramatically on the 
arrangement of holes and pairs around the FL.

\subsection{$N_G(I)$ formula}

We now turn to the evaluation of $N_G(I)$ for a general state.
One naively expects that this formula should be given by the
sum of pairs, $N_p$, and holes, $N_h$, above and below the FL 
respectively, i.e.\ $N_G(I) = N_p + N_h$. In fact $N_p=N_h$ 
since every pair above the FL comes from a hole below it. 
However this ansatz does not always work as we have already seen 
above. For example, according to this formula, 
the state $I_1$ of Fig.~\ref{fig1}b would have 2 instead of $N_G=1$, 
while the state $I_2$ of fig.~\ref{fig1}c has $N_G=2$, which 
is the correct value. 

Let us  introduce for convenience 
the occupation representation
of the states $I$, where a pair, a hole
and the FL 
are depicted as 
\mbox{$\bullet, \circ$ and $|$}
respectively. In the cases discussed
above we obtain: 
\mbox{$I_0 =  \bullet \cdots \bullet \bullet \bullet  
|  \circ \circ \circ \cdots \circ $},
\mbox{$I_1 = \bullet \cdots \bullet \bullet  \circ | 
\bullet \circ \circ \cdots \circ$}, and 
\mbox{$I_2 = \bullet \cdots \bullet  \circ \bullet | 
\circ \bullet \circ \cdots \circ$}.

We have found an algorithm to compute
$N_G(I)$. Given an integer $\ell \geq 0$, 
let us split $I$ into three disjoint sets, 
$I = A_\ell \cup B_\ell \cup C_\ell$, 
where $A_\ell$ contains the
lowest $M-\ell\;$ levels, $B_\ell$ the next $2 \ell$
levels and $C_\ell$ the remaining $N - M- \ell\;$ ones.
For $\ell =0$, the set $B_0$ is empty, and 
$A_0$ (resp.\ $C_0$) contains all the levels below
(resp.\ above) the FL, while for $\ell \geq 1$ the set
$B_\ell$ contains the nearest $\ell$ levels above and below
the FL. As an example, let us choose a state of the form
$I_3 = \bullet \stackrel{p}{\cdots} \bullet  
\circ \circ \bullet \bullet \bullet \bullet
| \circ \bullet \bullet \circ \stackrel{h}{\cdots} \circ$.  
For $\ell = 2$ the 
partition of $I_3$  is given by
$\{ \bullet \stackrel{p}{\cdots} \bullet  
\circ \circ \bullet \bullet \} \{  \bullet \bullet
| \circ   \bullet \} \{  \bullet \circ \stackrel{h}{\cdots} \circ \}$.  

Let us define the
number of pairs and holes for each set, i.e.\ 
$N_{p/h}^X ( X = A_\ell,B_\ell,C_\ell)$. The algorithm
giving $N_G(I)$ is  
\begin{eqnarray}
& N_G(I) =   \min_{\ell=0, \dots, 2 N_p}  N_G(I,\ell) & \label{5} \\
& N_G(I,\ell)  \equiv N^{A_\ell}_h + 
\min (N^{B_\ell}_h, N^{B_\ell}_p) + N^{C_\ell}_p \;. & 
\nonumber 
\end{eqnarray}
\no Applying this formula to $I_3$ one gets
$\{ N_G(I_3,\ell) \}_{\ell=0}^4 = \{ 4,5,4,3,4 \} $ and thus,
$N_G(I_3) = 3$. 
The value of $\ell_I$, which minimizes $N_G(I,\ell)$,
is given in this case by 3 (in general $\ell_I$ is not
equal to $N_G(I)$). 
The result of this formula is bounded,
$N_p \le N_G \le 2N_p$, and therefore any state 
with a finite $N_p$ would contain a finite
number of Gaudin energies.
The uncorrelated  
counting formula proposed earlier coincides
with the case $\ell_I=0$, since $N_G(I,0) = N_p + N_h$
(notice that $N_h = N^{A_0}_h$ and $N_p = N^{C_0}_p$).

The physical mechanism underlying Eq.~(\ref{5}) is 
the collective behavior of the holes and pairs
that occupy the $\ell_I$ closest levels to the FL. 
In a certain sense, $\ell_I$ measures the range
of correlation involved in the creation
of the elementary excitations out of
an initial pair-hole configuration. 
However, this correlation can be lifted introducing a shifted 
Fermi level $FL(\ell_I)$ defined by moving the FL an amount of 
$\ell_I$ levels downwards (resp.\ upwards) whenever $N_p^{B_\ell}$ 
is lower or equal (resp.\ greater) than $N_h^{B_\ell}$. 
This new Fermi level defines a new Fermi state out of which
the excited state with $N_G$ finite energies
is obtained by the creation of uncorrelated
pairs above and holes below the $FL(\ell_I)$. 
This construction provides a pathway to
the g.c.\ formulation as discussed later.

The formula presented here allows us to prove Gaudin's conjecture
by looking for all the states with a given $N_G$, and finding out
that its number corresponds to $d_{N_G}$ as stated in 
Table~\ref{table1}, the same way we already did with those with 
\mbox{$N_G = 1$ and $2$}. 


\subsection{Young diagrams representation}    
\label{Young}

The correlated behavior of the excitations is made 
more explicit by a pictorial representation of the states. 
The idea is to associate to every set $I$ 
a path $\gamma_I$ with $N$ links on the square lattice
${\cal Z}^2$, starting at the origin $(0,0)$.
This is achieved by associating a horizontal link
directed to the right, to every
hole $\circ$, and a vertical link directed 
upwards, to every pair $\bullet$. The map
starts from the lowest energy level and ends at 
the highest one. 
For example the path associated to the Fermi 
state 
\mbox{$I_0 = \bullet \stackrel{M}{\cdots}
\bullet \;|\; \circ \stackrel{N-M}{\cdots}  
\circ$} is a polygonal
line joining the points
\mbox{$(0,0) \rightarrow (0,M) \rightarrow (N-M,M)$}.
If $I$ describes a state with $M$ pairs
and $N$ energy levels, then the path
$\gamma_I$ ends at the point $(N-M,M)$. 
The number of these sort of paths
is $C^N_M$, which is precisely
the dimension of the Hilbert space ${\cal H}^N_M$
\cite{torres}.
 
\begin{figure}[t]
\begin{center}
\includegraphics[width= 8 cm,angle= 0]{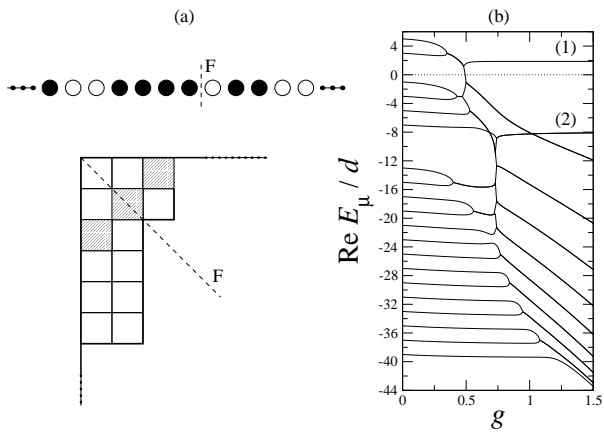}
\end{center}
\caption{a) The path and Young diagram of $I_3$. 
b) Real part of  $E_\mu$
for $I_3$. For $g$ large enough  
there is a real root (1) and a complex
root (2).}
\label{fig3}
\end{figure}

\begin{figure}[h]
\begin{center}
\includegraphics[width= 8 cm,angle= 0]{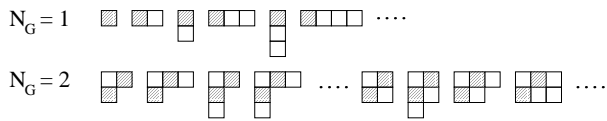}
\end{center}
\caption{YD's corresponding to $N_G=1$ and 2.} 
\label{fig4}
\end{figure}

In Fig.~\ref{fig3}a we depict the occupation and path representations 
of the state $I_3$ which yields $N_G= 3$, in agreement with 
the numerical results shown in Fig.~\ref{fig3}b. 
Moreover, Fig.~\ref{fig3}a illustrates the fact that any state $I$ 
gives rise to a Young diagram (YD) $Y_I$, whose boundary
is formed by the links which belong either to $\gamma_I$
or to $\gamma_{I_0}$, but not to both. 
The YD of the Fermi state is by construction empty, 
i.e.\ $Y_{I_0}$ = \O.  These YD's capture the basic properties of the
excitations. First of all, $N_G(I)$, given by Eq.~(\ref{5}), 
coincides with the number of squares on the longest \mbox{SW-NE} diagonal  
on $Y_I$ (see Fig.~\ref{fig3}a). 
This fact provides a geometrical meaning to $N_G(I)$ and leads to 
a combinatorial proof of Gaudin's conjecture, which can be stated
as follows: $d_{N_G}$ is the number of YD's, $Y_I$, associated to 
the paths $\gamma_I$, which have  $N_G$ squares on their longest 
SW-NE diagonal.  The proof of this conjecture uses the  methods of 
Ref.~\onlinecite{torres}\cite{Alex}. 
This result serves to classify the excitations in terms of YD's.
For example the states with $N_G=1$ and $2$ discussed above 
correspond to the YD's shown in Fig.~\ref{fig4}. 


Another properties of these diagrams 
are:  i) the pair-hole transformation  
of the states induces a 
transposition of their associated YD's,
ii) the main NW-SE diagonal on a YD coincides
with the FL (see Fig.~\ref{fig3}a),  
and  ii) the number of boxes of $Y_I$
is the excitation energy of $I$ 
(in units of $2d$) at $g=0$  
for the equally space model.


\section{Thermodynamic limit}
\label{Large-N}

As we explained in the previous Section, 
by increasing $g$, $M-N_G$ of the energies $E_\mu$ 
become complex and arrange themselves into an arc
which escape to infinity for large $g$, while 
the remaining $N_G$ stay finite with their positions
barely modified (see Figs.~\ref{fig1}b,c and \ref{fig5}).

Following the procedure presented in 
Refs.~\onlinecite{G-book,R-limit,large,amico} we take
the large $N$ limit keeping $M/N$, $g$ and $N_G$ finite.
In this limit the arc formed by the energies 
in the complex plain becomes dense, and allows for a 
continuous formulation.
In particular, the GS corresponds to an arc
$\Gamma_{I_0}$ in the complex energy plane, 
which in the $g \rightarrow \infty$ limit
goes to infinity. 

Excited states contain finite energies in addition
to the arc. A given finite root $E_\alpha$ 
can be either real or complex. In the former
case we shall call it a 1-string. 
In the latter case $E_\alpha^*$ is also a root,
which together with $E_\alpha$ form a 2-string
(an example of such states can be seen in Fig.~\ref{fig3}b).
There are also 3-strings formed by 
one real root and two complex ones, 
having approximately the same real part, and so on.
In general $\{ E_\alpha \}_{\alpha=1}^{N_G}$
is a combination of strings with 
several lengths. 
The remaining  $M - N_G$
roots condense into an arc $\Gamma_I$, which is a slight
perturbation of the GS arc $\Gamma_{I_0}$. 
In Fig.~\ref{fig5}
we depict $\Gamma_{I_0}$ and $\Gamma_{I}$ 
for the two excited states $I_1$ and $I_2$
shown in Figs.~\ref{fig1}b,c.

Taking into account these considerations, and
using the methods of Refs.~\onlinecite{G-book,large} 
one can show in the large $N$ limit that  
the excitation energy of a Richardson state $I$ is given by
\beq
E_{exc} = \sum_{\alpha =1}^{N_G} 
\sqrt{ (E_\alpha - \ve_0)^2 + \Delta^2} \;,
\label{6}
\eeq
\no where $\ve_0$ is twice the chemical potential,
and the energies $E_\alpha$ satisfy the modified Gaudin equations
\beq 
0= \sum_{j=1}^N \frac{1}{ R(\ve_j) (\ve_j - E_\alpha)}
- \sum_{\beta \neq \alpha}^{N_G} 
\frac{2}{ R(E_\beta)( E_\beta - E_\alpha)} \;, 
\label{7}
\eeq
\no with $R(E) = \sqrt{ (E - \ve_0)^2 + \Delta^2}$.
As $g \to \infty$ one has $\Delta \sim g \omega$ and 
Eqs.~(\ref{7}) become Eqs.~(\ref{4}). 

\begin{figure}[t]
\begin{center}
\includegraphics[width= 6 cm]{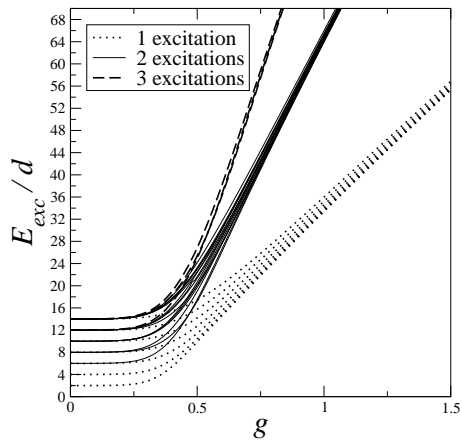}
\end{center}
\caption{Excitation energies
$E_{exc}=E - E_{GS} \leq 14 d$ 
for $M= 20$ pairs at half filling.
There are 44 =13+26+5 states 
corresponding to $N_G= 1,2,3$
respectively. The particle-hole symmetry
reduces these numbers
to 25 = 7+15+3. }
\label{fig2}
\end{figure}

The excitation energy given by Eq.~(\ref{6}) 
fits quite well the excitation energies of our prototype
example ($N = 40$, $M = 20$), as shown in Fig.~\ref{fig2}.
This also exhibits the linear behavior of the excitation
energy for $g \to \infty$, i.e.\ $E_{exc} \sim N_G \Delta$ 
as stated in Section~\ref{Richardson}, and in full
agreement with the large $g$ behavior of Eq.~(\ref{6}).
Thus we can extend our conjecture to the whole range of $g$.
Namely, any excited state is composed by $N_G$ elementary 
excitations associated to the finite Gaudin energies.
The Hilbert space spanned by these excitations has therefore
a dimension $d_{N_G} = {N\choose N_G} - {N\choose N_G-1}$.  
Hence, it is reasonable to call this new
type of excitations Gaudin pairs or {\em gaudinos}.

In order to compare our results with the BCS standard solution
lets consider the excitation energy given by a {\em real Cooper
pair} in the Bogoliubov approach, which is given by
$\sqrt{\ve_j^2 + \Delta^2}$ (notice that $\Delta \equiv 2\Delta_{BCS}$),
and span a Hilbert space of dimension ${N \choose N_G}$.
The standard Bogoliubov quasiparticle with an energy 
$\sqrt{\ve_j^2 + \Delta^2}/2$ would have to be compared 
with excitations involving broken Cooper pairs.
Since $E_{\alpha}$ in Eq.~(\ref{6}) lies between two energy levels, with
$\ve_{j+1} - \ve_j = 2d \sim 1/N$ (e.g.\ in Fig.~\ref{fig1}b 
$E_{20}(\infty) = 0$ with $\ve_{20} < E_{20} < \ve_{21}$),
$E_{\alpha} = \ve_j + O(1/N)$, 
and $d_{N_G} = {N \choose N_G}[1 - \frac{N_G}{N-N_G+1}]$.
Therefore, our theory is consistent within $O(1/N)$ 
corrections, as it is well known from the existing relation 
between a canonical and a grand canonical ensemble formulation
in the statistical physics.


It is important to notice how the Bogoliubov excitations are
uncorrelated with respect to the BCS ground state.  We already 
pointed out that the correlation present in our formulation is 
lifted by choosing a shifted Fermi level $FL(\ell_I)$.  This Fermi level 
is within a distance $O(1/N)$ from the original one.
The selection of a new Fermi level leads a new Fermi sea 
(with a different number of particles), allowing for a
grand canonical formulation in a natural way.

In summary, our {\em gaudinos} will yield the
same results as the BCS theory in the extrict $N = \infty$ limit, 
and will account for the exact corrections to the bulk results
for the finite size superconducting grains for all the physical
observables and thermodynamic properties.


\section{Conclusions}
\label{Conclusions}

We have shown in this paper that the elementary excitations 
of the exactly solvable BCS model in the canonical ensemble 
can be explained by the Gaudin model and have no counterpart
in the Bogoliubov picture of quasiparticles.
Their peculiar dispersion relation and the unexpected 
counting properties, which are due to the correlated behavior
of pairs and holes around the Fermi level, provide the exact 
finite size corrections to the BCS bulk results, valid for
large systems in the g.c.\ ensemble.
These excitations, together with those obtained by breaking 
Cooper pairs, supply the complete spectrum of the canonical 
BCS model.
A formula to compute the number of elementary excitations
for any given state was also proposed.

We explained how the description in terms of gaudinos
agrees with the Bogoliubov picture in the thermodynamic 
limit to leading order in $N$.  In the case of broken pairs, 
which was not presented here, the mechanism is identical.
It is of interest to study how the phase of the superconducting
order parameter emerges from this fixed number of particles
formulation.  It will be intimatelly related to the possibility
of choosing a shifted Fermi level in the large $N$ limit
(which looses the correlation of the excitation), 
allowing the introduction of ground states with different
number of pairs.

Although we used as an example a system of equally
spaced levels, the results are more general, and apply
to any distribution of levels.  This assertion is
based on numerical calculations considering
broken Cooper pairs. In this case bloked levels are removed, 
and we are left with a non-equally spaced spectrum,
obtaining again the same general results.
In the case of a non-constant pairing we also expect the 
qualitative picture presented here to hold. 


\section*{Acknowledgments}

We thank R.\ W.\ Richardson, C.\ Essebag, A.\ Di Lorenzo,
A.\ Mastellone, L.\ Amico and A.\ Berkovich for discussions. 
This work has been supported by the grants 
BFM2000-1320-C02-01/02. 




\begin{thebibliography}{99}

\bibitem{BCS} J. Bardeen, L.N. Cooper and J.R. Schrieffer,
Phys. Rev. {\bf 108}, 1175 (1957).

\bibitem{nuclear} A. Bhor, B. Mottelson and D. Pines,
Phys. Rev. {\bf 110}, 936 (1958). 

\bibitem{DMRG} J. Dukelsky and G. Sierra, Phys. Rev. Lett. {\bf 83},
172 (1999); Phys. Rev. {\bf B 61}, 12302 (2000).

\bibitem{exact} R.W. Richardson, Phys. Lett. {\bf 3}, 277 (1963);
R.W. Richardson,  N. Sherman,  Nucl. Phys. {\bf B 52},
(1964) 221. 

\bibitem{random} G. Sierra, J. Dukelsky, G. G. Dussel, 
J. von Delft and  F. Braun,
Phys. Rev. {\bf B61}, 11890 (2000).

\bibitem{vDR} J. von Delft, D.C. Ralph,
Phys. Reps.,{\bf 345}, 61 (2001).

\bibitem{G-book} M. Gaudin, 
``Mod\`eles exactament r\'esolus'', Les \'Editions
de Physique, France, 1995. 

\bibitem{torres} F.M. Goodman, P. de la Harpe, V.F.R. Jones,
``Coxeter Graphs and Towers of Algebras'', Springer-Verlag,
New York, 1989. 

\bibitem{Alex} We thank A.\ Berkovich for providing us with an
alternative proof based on counting RSOS paths.

\bibitem{R-limit} R.W. Richardson,
J. Math. Phys. {\bf 18}, 1802 (1977).

\bibitem{large} J.M. Rom\'an, G. Sierra, J. Dukelsky,
Nucl. Phys. {\bf B634} [FS] (2002) 483. 

\bibitem{amico} L. Amico, A. Di Lorenzo, 
A. Mastellone, A. Osterloh, R. Raimondi, 
Annals of Physics vol. 299, 228 (2002). 

\end{thebibliography}
\end{document}